\documentstyle[aps,prb,multicol,epsf
]{revtex}
\def\Cfrac{\mathop{\rm K}}
\def\nb{\mbox{\boldmath $\nabla$}}

\def\ie{{\em i.e.}}
\def\re{\Re{\rm e}}

\def\cfrac#1#2{\frac{\displaystyle #1}{\displaystyle #2\!}}

\def\wid{\end{multicols}
\widetext
\noindent\rule{20.5pc}{0.1mm}\rule{0.1mm}{1.5mm}\hfill}
\def\nar{\hfill\rule[-1.5mm]{0.1mm}{1.5mm}\rule{20.5pc}{0.1mm}
\begin{multicols}{2}
\narrowtext}

\begin{document}
\advance\textheight by -.20in

\title{Single electron magneto-conductivity of a nondegenerate
  2D electron system in a quantizing magnetic field}

\author{Frank~Kuehnel$^{1}$, Leonid P.~Pryadko$^{2,3}$, and
M.I. Dykman$^1$}

\address{$^{1}$Department of Physics and Astronomy, Michigan State
  University, East Lansing, Michigan, 48823}

\address{$^2$School of Natural Sciences, Institute for Advanced Study,
  Princeton, New Jersey, 08540}
\address{$^3$Department of Physics, University of California, Riverside,
California, 92521}

\date\today
\maketitle

\begin{abstract}
  We study transport properties of a non-degenerate two-dimensional
  system of non-interacting electrons in the presence of a quantizing
  magnetic field and a short-range disorder potential.  We show that
  the low-frequency magnetoconductivity displays a strongly asymmetric
  peak at a nonzero frequency.  The shape of the peak is restored from the
  calculated $14$ spectral moments, the asymptotic form of its
  high-frequency tail, and the scaling behavior of the conductivity
  for $\omega\to 0$. We also calculate $10$ spectral moments of the
  cyclotron resonance absorption peak and restore the corresponding
  (non-singular) frequency dependence using the continuous fraction
  expansion.
  Both expansions converge rapidly with increasing number of included
  moments, and give numerically accurate results throughout the region
  of interest. We discuss the possibility of experimental observation
  of the predicted effects for electrons on helium.
\end{abstract}

\begin{multicols}{2}
\narrowtext

\section{Introduction}

Single-electron dynamics in the lowest Landau level (LLL) broadened by
a delta-correlated scalar disorder potential provides the simplest
framework for the analysis of the integer quantum Hall effect (IQHE).
Wegner's exact calculation \cite{Wegner-83} of the density of states
raised hopes that the corresponding model may be exactly solvable, and
much effort was put into understanding transport in this model.
However, in contrast to the density of states, the conductivity is
expressed in terms of a two-particle Green's function and depends not
only on the energies of single-particle states, but also on their wave
functions.

A noteworthy feature of the single-electron wave functions in a strong
magnetic field is delocalization near the centers of the
disorder-broadened Landau bands. For small energies $E$ counted off
from the band center (in the neglect of band mixing), the localization
length $\xi$ diverges as a universal
power\cite{Chalker-Coddington,Huckestein-95} of 
$|E|$.  These are transitions between large-radius states that
form the low-frequency conductivity of the system.  As a result, for
Fermi energies close to a band center, the model displays a universal
critical behavior at sufficiently small temperatures and frequencies.

The width of the critical region depends on the properties of the
disorder potential and the Landau level number.  For the lowest Landau
level, it is of the order of the bandwidth $\hbar\gamma$, the only
dimensional parameter of the Hamiltonian projected on the LLL (we
assume that the cyclotron frequency $\omega_c\gg \gamma$).  Outside
the scaling region, the spatial extent of the eigenstates is small, of
the order of the magnetic length $l= (\hbar/m\omega_c)^{1/2}$, and the
universality is lost. Therefore the overall frequency dependence of
the conductivity is determined by the disorder mechanism and may allow
to 
discriminate between different mechanisms.  For this reason, it is
interesting to obtain the frequency dependence of the conductivity,
including its universal and non-universal parts, at least for some
basic models of disorder. It is also of interest to find highly
accurate numerical results, as they may be used to test various
approximate analytical approaches.

In the present paper we consider the frequency-dependent
conductivity for a short-range disorder potential.
Short-range disorder is usually a good model of the random
potential experienced by electrons trapped on the surface of liquid
helium. This potential is due to quasistatic ripplons, slow helium
vapor atoms, or, for thin helium films, substrate disorder
\cite{Andrei-Book}. The electron correlations in this system are
usually strong. However, it was believed that, for strong
enough magnetic fields, the static magnetoconductivity can be
described in the single-electron approximation (for a review see
Refs.~\CITE{Lea-in-Andrei,Lea-98,Teske-99}).  The corresponding results
obtained within the self-consistent Born approximation\cite{Ando-82}
(SCBA) appeared to be in reasonable agreement with the
experiment\cite{Adams-88B,Heijden-88,Scheuzger-94}.

The SCBA ignores the interference effects which lead to electron
localization and to the scaling behavior near the band
center. Therefore there is an apparent contradiction between the
interpretation of the experimental data on electrons on helium and the
phenomenology of the quantum Hall effect. From this point of view, it
is important to develop a single-electron theory of magnetotransport
that will take localization effects into account and will thus extend
the ideas of the IQHE theory to a new parameter range, which is of
interest for electrons on helium in particular. 
The results can serve
as a basis for the full many-electron theory of magnetoconductivity at
strong magnetic fields\cite{Kuehnel-00}.

With this in mind, we will analyze magnetotransport in the regime
\begin{equation}
  \label{eq:non-degeneracy}
  \hbar\gamma\ll k_{B} T, \quad nl^2\ll1, 
\end{equation}
where the disorder-induced broadening of the lowest Landau level is
small compared to temperature, and simultaneously the filling fraction
is small, \ie, the electron system is nondegenerate.  Much of the
experimental data on electrons on helium refers to the range
(\ref{eq:non-degeneracy}), as the system is very clean, and the
experiments are often done at low electron densities $n\sim
10^8$~cm$^{-2}$. This range is also of interest for low-density
electron systems in semiconductors.

In the range (\ref{eq:non-degeneracy}) all states within a broadened
Landau level are nearly equally populated, and one no longer needs to
take into account the Boltzmann factor while computing temperature and
disorder averages using the Hamiltonian projected on that level. As we
show below, this simplification allows us to make accurate
calculations of the frequency-dependent conductivity using the method
of moments, which was previously suggested for this problem by one of
us\cite{Dykman-78}. 

\begin{figure}[htb]
  \begin{center} \epsfxsize=\columnwidth \epsfbox{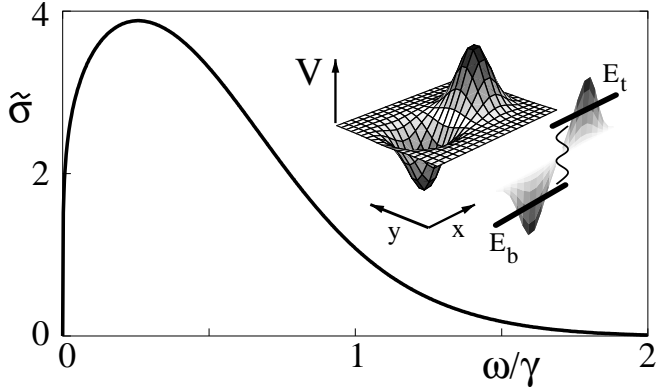}
    \end{center} \caption{Reduced microwave con\-duc\-tiv\-i\-ty
    (\protect\ref{eq:kubo-formula2}) of a non-inter\-act\-ing 2DES at
    the lowest Landau level in a short-range disorder potential for
    $\omega, \gamma\ll k_BT/\hbar$.  For small frequencies, $\omega\ll
    \gamma$, the conductivity is singular, $\sigma_{xx}\sim
    \omega^\mu$. It is determined by spatially large, nearly
    delocalized states.  For $\omega\gg \gamma$, the conductivity is
    determined by large optimal fluctuations of the disorder potential
    as illustrated in the inset.  The optimal potential $V_{\rm
    opt}({\bf r})$ is such that $\hbar\omega$ equals to the energy
    difference $E_{\rm t}-E_{\rm b}$ between the top and bottom bound
    states $|t\rangle, \,|b\rangle$, and at the same time these states
    are maximally overlapping.}  \label{fig:conduct}
\end{figure}

\begin{figure}[htb]
  \begin{center}
    \epsfxsize=\columnwidth
    \epsfbox{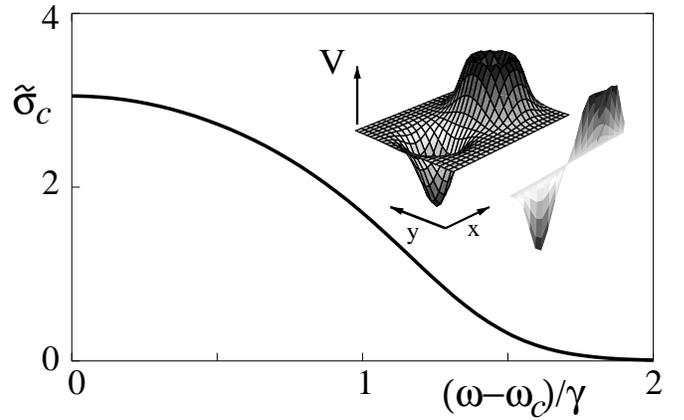}
  \end{center}
  \caption{Reduced conductivity at the cyclotron absorption peak [see
    Eq.~(\protect\ref{red_sigma_cyc})] of a non-inter\-act\-ing 2DES
    in a short-range disorder potential for $|\omega-\omega_c|, \gamma\ll
    k_BT/\hbar\ll\omega_c$.  The curve is analytic in the center of
    the peak.  The absorption at the tails, $|\omega-\omega_c|\gg
    \gamma$, is determined by large optimal fluctuations of the
    disorder potential as illustrated in the inset.}
    \label{fig:cyc_conduct}
\end{figure}

The outline of the paper is as follows. In Sec.~II we calculate the
frequency-dependent conductivity $\sigma_{xx}(\omega)$ in the
temperature range $\hbar\gamma\ll k_BT$, $\exp(\hbar\omega_c/k_BT)\gg
1$ for $\omega\ll \omega_c$ [these results were previously announced
in Ref.~\CITE{Kuehnel-00}].  We find the asymptotics of the conductivity
at both small ($\omega\to0$) and comparatively large
($\gamma\ll\omega\ll\omega_c$) frequencies, and show that
$\sigma_{xx}(\omega)$ has a peak at a nonzero frequency
$\omega\sim\gamma$.  Using an efficient diagram classification scheme,
we compute exactly the first $14$ spectral moments of this peak.
These moments contain information about the short-time
($\sim\gamma^{-1}$) dynamics of the system.  Combined with the low and
high-frequency asymptotics, they allow us to accurately restore the
entire function $\sigma_{xx}(\omega)$ [see Fig.~\ref{fig:conduct}].
In Sec.~III we investigate the cyclotron resonance,
\ie, $\sigma_{xx}(\omega)$ for $\omega\approx \omega_c$. We calculate
the first $10$ frequency moments of the cyclotron resonance absorption
peak and use them to accurately restore its shape [see
Fig.~\ref{fig:cyc_conduct}] as a function of frequency detuning
$\Delta\omega\equiv \omega-\omega_c$. To do the restoration, we also
calculate the asymptotic form of the tails of the cyclotron resonance
using the method of optimal fluctuation [our result differs from that
obtained earlier by Ioffe and Larkin\cite{Ioffe-81}]. In Sec.~IV we
discuss the ways to observe the predicted here behavior in experiment.
Technical details are given in the Appendices.

\section{Low-frequency single-electron conductivity}
\label{sec:low}

In this section, we calculate the conductivity of a nondegenerate
non-interacting
two-dimensional electron system (2DES) for low frequencies, $\omega\ll
k_BT/\hbar$. We consider the case of a delta-correlated disorder
potential and assume that the broadening of the Landau level
$\gamma\ll \omega_c,\,k_BT/\hbar$ [cf.\ Eq.~(\ref{eq:non-degeneracy})].

In the range (\ref{eq:non-degeneracy}) all states within the LLL
are equally occupied, and the Kubo formula for the dissipative
conductivity can be written as a simple trace without the Boltzmann
factor,
\begin{eqnarray}
  \!\sigma_{xx}(\omega)&=&{n\left(1-e^{-\beta
  \omega}\right)\over\hbar\omega} \, \re\int_0^{\infty} \! dt\,
  e^{i\omega t}\, \bigl\langle{j}_x(t)\,{j}_x(0)\,
  \bigr\rangle\quad\strut \label{eq:kubo-full}\\
  \label{eq:kubo-simplified} &\approx&
  {n\beta\over2\hbar}\!\int_{-\infty}^\infty \!\!\! dt\, e^{i\omega
  t}\, {\rm Tr}_0\overline{\bigl\{j_x(t)\,j_x(0)\bigr\}},\quad
  \beta\omega\ll1.
\end{eqnarray}
Here $j_x\equiv ep_x/m$ is the one-electron current operator,
$\beta\equiv\hbar/k_BT$, the angular brackets $\langle\cdot\rangle$
denote statistical averaging over the states followed by an averaging
over quenched disorder, while the horizontal line denotes only the
disorder averaging.  The trace Tr$_0$ in
Eq.~(\ref{eq:kubo-simplified}) is performed over all single-particle
states of the lowest Landau level; the energies are measured with
respect to its center. Eq.~(\ref{eq:kubo-full}) is written for the
case of strongly quantizing magnetic fields, $\exp(\beta\omega_c)\gg
1$, so that only the lowest Landau level is occupied. However, the
calculation is readily generalized to the case of arbitrary
$\beta\omega_c$ by replacing Tr$_0$ by the sum of traces over the
states of each Landau level $n$ weighted with
$\exp(-n\beta\omega_c)[1-\exp(-\beta\omega_c)]$.

Calculations within a single Landau level are conveniently done using
the formalism of the guiding center coordinates ${\bf R}\equiv (X,Y)$.
The electron dynamics in the random potential $V({\bf r})$ is mapped
onto that of a 1D quantum particle with the generalized momentum and
coordinate $X$ and $Y$, and with the Hamiltonian
\begin{equation}
  \label{eq:hamiltonian}
  H=\hbar\gamma \sum\nolimits_{\bf q} \tilde V_{\bf q}\,\exp(i{\bf
  qR}),\quad [X,Y]=-il^2.
\end{equation}
Respectively,
 the guiding center velocity
is determined 
by the potential gradient,
\begin{equation}
  \label{eq:guiding-velocity}
  \dot{R}_\mu=-il^2\gamma\sum_{\bf q}
  \epsilon_{\mu\nu}\,{q}_\nu \,  \tilde
  V_{\bf q}\,e^{i{\bf q}{\bf R}}, 
\end{equation}
where $\mu,\nu=x,y$, and $\epsilon_{\mu\nu}$ is the unit antisymmetric
tensor, $\epsilon_{xy}=-\epsilon_{yx}=1$. 

The dimensionless coefficients 
\begin{equation}
  \tilde{V}_{{\bf q}}\equiv
  (V_{\bf q}/\hbar\gamma)\exp(-l^2q^2/4)
  \label{eq:effective-potential-lll}
\end{equation}
are proportional
to the Fourier components of the disorder potential,
\begin{equation}
  V_{\bf q}\equiv S^{-1}\int
  d^2{\bf r}\,V({\bf r})e^{-i{\bf qr}},\label{eq:fourrier-potential}
\end{equation}
where $S$ is the area of the system.  
For higher Landau levels the coefficients $\tilde V_{\bf q}$ have to
be modified as explained in Appendix~\ref{app:guiding} [see
Eq.~(\ref{eq:projected-potential})].  We 
will assume that $V({\bf r})$ is zero-mean Gaussian and delta-correlated,
\begin{equation}
  \label{potential}
  \overline{ V({\bf r})\,V({\bf r}')} = v^2\delta({\bf
    r}-{\bf r}'),
\end{equation}
in which case the SCBA width of the lowest Landau band is\cite{Ando-82}
$\hbar\gamma = (2/\pi)^{1/2}v/ l$.

In the simplified Kubo formula~(\ref{eq:kubo-simplified}) the
temperature dependence is factorized, and we can rewrite the
low-frequency conductivity in the form of the generalized Einstein
relation
\begin{equation}
  \label{eq:einstein-relation}
  \sigma_{xx}(\omega)={ne^2D\over k_BT}
  \,{1\over8}\,\tilde\sigma(\omega), 
\end{equation}
where $D=l^2\gamma$ is the characteristic diffusion coefficient and,
as discussed in Appendix~\ref{app:guiding}, 
\begin{equation}
  \label{eq:kubo-guiding}
  \tilde\sigma(\omega)\equiv
  \frac{2}{l^2\gamma}\int_{-\infty}^{\infty}dt\,e^{i\omega t}
  {\rm Tr}_0
    \Bigl\{\overline{\dot{\bf R}(t)\cdot\dot{\bf R}(0)
  }\Bigr\}
\end{equation}
is the reduced conductivity. It depends on the ratio $\omega/\gamma$
of the only two quantities with the dimension of frequency that remain
after projection on one Landau level.

The expression~(\ref{eq:kubo-guiding}) can be rewritten
with the help of Eq.~(\ref{eq:guiding-velocity}) as
\begin{eqnarray}
  \label{eq:kubo-formula2}
  \tilde\sigma(\omega)=&& 
  -{2l^2\gamma}\int_{-\infty}^{\infty}dt\,e^{i\omega t}
  \sum\nolimits_{{\bf q},{\bf q}'}\left({\bf q}\,{\bf q}'\right)
  \nonumber \\  
  &&\times {\rm Tr}_0\Bigl\{ \overline{\tilde V_{\bf q}\tilde
  V_{{\bf q}'}\exp\left[i{\bf q\,R}(t)\right]
  \exp\left[i{\bf q}'\,{\bf R}(0)\right]}
  \Bigr\}. 
\end{eqnarray}
This form is particularly convenient for calculating 
the frequency moments of the reduced conductivity, 
see below in Sec.~\ref{subsec:moments}.

Yet another representation of the reduced low-frequency conductivity
can be obtained if we describe time evolution of the electron
operators in Eq.~(\ref{eq:kubo-formula2}) using the set $|n\rangle$ of the
eigenstates of the full electron Hamiltonian for the lowest Landau
level and perform the time integration,
\begin{equation}
  \label{eq:kubo-gradient}
  \tilde\sigma(\omega)={4\pi l^2\over\hbar\gamma}
  \sum_{n,m}\overline{\delta(E_n-E_m-\hbar\omega) \left|\bigl\langle
  n\bigl|\nb V\bigr|m\bigr\rangle\right|^2},    
\end{equation}
where $E_n$ are the energies of the LLL states $|n\rangle$ in the
potential $V({\bf r})$ (again, generalization to the case of several
occupied Landau levels is straightforward).

We emphasize that, in the chosen parameter range, the Landau-level
projection resulted in expressions that {\em do not\/} contain the usual
disorder-dependent denominator, and the quenched disorder averaging
can be done directly, without invoking supersymmetry or the replica trick.

\subsection{Tail of the low-frequency conductivity}
\label{sec:low-tails}

We begin with calculating the asymptotic form of the reduced
conductivity $\tilde\sigma(\omega)$ for $\omega\gg\gamma$ from
Eq.~(\ref{eq:kubo-gradient}). In the neglect of inter-band mixing, the
energies $E_n$ are symmetrically distributed around the Landau band
center ($E=0$).  The tails of the density of states $\rho(E)$ are
known to be Gaussian, $\rho(E) \propto \exp(-4E^2/\hbar^2\gamma^2)$.
They are determined by the probability of the optimal (least
improbable) potential fluctuation $V_E({\bf r})$ in which the lowest
or highest bound state has energy $E$
($|E|\gg\gamma$)\cite{Ioffe-81,Wegner-83,Benedict-87}.

If we ignore the matrix element in Eq.~(\ref{eq:kubo-gradient})
altogether (as we show below, this only affects the prefactor), the
tail of the conductivity will be proportional to the probability to
find two states $E_n$, $E_m$ such that $E_n-E_m=\hbar\omega$.  The
major contribution comes from states at the opposite ends of the
energy band with energies close to $E_n=-E_m=\hbar\omega/2$, giving
\begin{equation}
  \label{eq:large-omega}
  \tilde\sigma(\omega) \propto [\rho(\hbar \omega/2)]^2
\propto \exp(-2\omega^2/\gamma^2).
\end{equation}

To check this approximation, we will apply the method of optimal
fluctuation\cite{Halperin-Lax-66,Ioffe-81}.  The averaging over
disorder in Eq.~(\ref{eq:kubo-gradient}) will be done using the path
integral representation
\begin{equation}
  \label{eq:path-integral}
  \overline{{\cal F}[V]} \equiv \int {\cal D} V({\bf r})\,{\cal F}[V({\bf
    r})]\exp\{-{\cal R}[V({\bf r})]\},
\end{equation}
where, for a delta-correlated Gaussian potential with the
correlator~(\ref{potential}), 
\begin{equation}
  \label{eq:weight}
  {\cal R}[V] = {1\over 2 v^2}\int  d{\bf r}\, V^2({\bf r}) .
\end{equation}

For large $\omega$,
the leading contribution to the sum (\ref{eq:kubo-gradient}) comes
from transitions between the states $|\psi_{\rm t}\rangle$ and
$|\psi_{\rm b}\rangle$ with energies $E_{\rm t}$ and $E_{\rm b}$ at
the top and bottom of the Landau band, respectively,
\begin{equation}
  E_{\rm t,b}=\int d{\bf r}\,V({\bf r})\,|\psi_{\rm t,b}({\bf r})|^2.
    \label{eq:variational-energies}
\end{equation}
To logarithmic accuracy, 
the conductivity is given by the solution of the variational problem
of finding the optimal potential $V({\bf r})$ which
minimizes the functional ${\cal R}[V]$ and maximizes the matrix
element of the transition subject to the constraint $E_{\rm t}-E_{\rm
b}=\hbar\omega$, \ie,
\begin{eqnarray}
  \label{eq:variational-sigma-two}  
    \tilde{\sigma}(\omega) &\propto& \max\limits_V \; \Bigl\{
 \exp  \left[-{\cal R}[V]+\lambda\,(E_{\rm t}-E_{\rm
 b}-\hbar\omega) \right] \nonumber \\
  & \relax &\times |\langle\psi_{\rm t} | \nb V | 
  \psi_{\rm b} \rangle |^2 \Bigr\}, 
\end{eqnarray}
where $\lambda$ is a Lagrange multiplier.
Variation with respect to $V({\bf r})$ gives the equation
\begin{equation}
  \label{eq:optimal} {V({\bf r}) \over v^2} = \lambda \left(
  |\psi_{\rm t}|^2 - |\psi_{\rm b}|^2 \right) + 
  {\delta\over \delta V({\bf r})}\ln|\langle\psi_{\rm t} | \nb V | 
  \psi_{\rm b} \rangle |^2
\end{equation}
(for brevity, we do not give the explicit form of the last term).

We have analyzed the variational problem using a simple and tractable
direct variational method, and also by finding the maximum in
Eq.~(\ref{eq:variational-sigma-two}) numerically. To see the
qualitative features of the solution, we first discuss it ignoring the
contribution of the matrix element. In this case the Lagrange
multiplier $\lambda$ is given by the consistency equation,
\begin{equation}
  \label{eq:variational-consistency}
  \hbar\omega = E_{\rm t} - E_{\rm b}
  = v^2 \lambda \int   d{\bf r}\,\left (|\psi_{\rm t}|^2 - |\psi_{\rm
  b}|^2\right)^2,  
\end{equation}
and then the conductivity~(\ref{eq:variational-sigma-two}) is
\begin{equation}
  \label{eq:variational-conductivity}
   |\ln  \tilde\sigma(\omega)|= \hbar^2\omega^2 \left[ 2v^2\!\int \!
    d{\bf r}\,\left (|\psi_{\rm t}|^2 - |\psi_{\rm
        b}|^2\right)^2\right]^{-1}. 
\end{equation}

The solution (\ref{eq:optimal}) corresponds to a potential of the form
of a well and a hump, far away from each other (cf.\
Fig.~\ref{fig:conduct}). The potential is antisymmetric, the well and
the hump have the same Gaussian shape [$\propto\exp(-r^2/2l^2)$, with
${\bf r}$ counted off from the corresponding extremum] and opposite
signs.  The wave functions $\psi_t$ and $\psi_b$ are localized at the
hump and the well of $V({\bf r})$, respectively, and are given just by
the most ``localized'' wave function of the lowest Landau level,
namely that with zero angular momentum, $\psi_{00}({\bf r})\propto
\exp(-r^2/4l^2)$, centered at the appropriate potential extremum.  The
overlap of these wave functions is negligibly small, and
Eqns.~(\ref{eq:variational-consistency}),
(\ref{eq:variational-conductivity}) give
$$
\hbar\omega=2v^2\lambda\, A,\quad
   |\ln  \tilde\sigma(\omega)|= {\hbar^2\omega^2\over
     4v^2\,A}={\omega^2\over 2\pi\,\gamma^2 l^2\,A},
$$
 with $A \equiv \int d{\bf
r}\,|\psi_{\rm t,b}|^4 =(4\pi l^2)^{-1}$. In this way we recover the
expression (\ref{eq:large-omega}) for the conductivity tail. For
higher Landau levels ($N\geq 1$), the wave functions have
the form\cite{Benedict-87} $\psi_{\rm t,b} \propto
r^N\exp(-iN\phi)\exp(-r^2/4l^2)$, in which case the corresponding
constant $A_N=(4\pi l^2)^{-1}(2N)!/2^{2N}(N!)^2$.

The prefactor in Eq.~(\ref{eq:variational-sigma-two}) prevents the
well and the hump of $V({\bf r})$ from being too far away from each
other.  Nevertheless, the full variational equation (\ref{eq:optimal})
has a solution with an antisymmetric 
optimal potential $V({\bf r})=-V(-{\bf r})$ and symmetric wave
functions $\psi_t({\bf r}) = \psi_b(-{\bf r})$; respectively,
$E_t=-E_b=\hbar\omega/2$.  To estimate the role of the overlap integral
we used the direct variational method in which we sought the potential
in the
form $V({\bf r}) = \tilde V(|{\bf r-r}_0|)-\tilde V(|{\bf r +
r}_0|)$ with $\tilde V(r)=V_0\exp(-r^2/2l^2)$. The distance  $2r_0$
separating 
the hump and the well was used as a variational parameter.
Given the potential, one has to solve the Schr\"odinger equation, 
looking for the wave functions projected on the lowest Landau level.
We took the functions $\psi_{\rm t,b}$ in the simplest form of
orthogonal combinations of the zero-momentum wave functions centered
close to $\pm {\bf r}_0$ (the positions were found using a variational
procedure).  The distance $r_0$ scales with frequency
logarithmically.  The overall asymptotic expression for the exponent in
$\tilde\sigma$ was the same as in Eq.~(\ref{eq:large-omega}); the overlap
integral gave only a prefactor,
$$ |\langle\psi_{\rm
  t} | \nb V | \psi_{\rm b}
\rangle |^2\sim (\hbar\gamma^2/l\omega)^2 \ln
(\omega/\gamma).
$$
[An extra $\omega$-dependent contribution to the overall prefactor in
$\tilde\sigma$ comes from the prefactor in the path integral
(\ref{eq:path-integral}). It actually increases with the increasing
$\omega$. However, an evaluation of this prefactor goes beyond the
scope of this paper, and in some sense is superseded by the results
obtained below with the method of moments.]

To further check the accuracy of the asymptotic behavior of
$\tilde\sigma(\omega)$, we maximized\cite{Kuehnel-thesis} the functional in
Eq.~(\ref{eq:variational-sigma-two}) numerically. 
We used the variational
equation (\ref{eq:optimal}) to represent the optimal potential as  
a bilinear
combination of the LLL wave functions $\psi_{0m}({\bf r}) \propto
r^m\exp(im\phi)\exp(-r^2/4l^2)$ with different magnetic
quantum numbers $m \geq 0$,
$$
V({\bf r})=\sum_{m,m'}\,u_{mm'}\,\psi^*_{0m}({\bf r})\,\psi_{0m'}({\bf
  r}). 
$$
The corresponding 
eigenfunctions $\psi_{\rm t,b}$ 
were written as linear combinations of the
same functions $\psi_{0m}({\bf r})$.  

\begin{figure}[htb]
  \begin{center}\strut\hfill%
    \epsfxsize=0.56\columnwidth
    \epsfbox{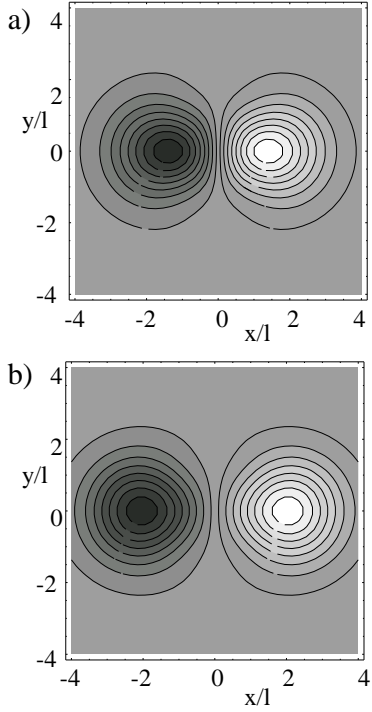}\hfill\strut
  \end{center}
  \caption{Density plot of the optimal potential for $\omega = 3\gamma$
    (a) and $\omega = 8\gamma$ (b).  The distances are measured in
    units of the magnetic length $l$.}
  \label{fig:optV}
\end{figure}

Both the exponent and the prefactor of the variational
functional~(\ref{eq:variational-sigma-two}) calculated numerically
become close to the result of the direct variational method
for $\omega/\gamma \agt 3$.   The shape of the optimal potential
found numerically for two values of $\omega/\gamma$ is illustrated in
Fig.~\ref{fig:optV}.
 
\subsection{Conductivity at small frequencies.}
\label{sec:small-omega}

An entirely different set of states defines the conductivity for very
small frequencies, $\omega\ll\gamma$.  In this regime the constraint
imposed by conservation of energy is not very restrictive, and it is
the matrix element that determines relative contributions of different
pairs of states.  
Close to the static
limit the contribution to the conductivity increases dramatically with
the typical size of a wavefunction.  As a result, $\tilde\sigma$
[Eq.~(\ref{eq:kubo-simplified})]  is primarily determined by a narrow energy
interval at the center of the Landau band where the states are nearly
delocalized. The energy of the band center ($E=0$) is a critical
energy, similar to the critical value of the control parameter in the
theory of classical percolation transition.  At small deviations from
the critical energy (parametrized by dimensionless energy
$\varepsilon\equiv E/\hbar\gamma$) the correlation length diverges,
$\xi_\varepsilon\sim l\, |\varepsilon|^{-\nu}$, where
$\nu=2.33\pm0.03$ is the localization
exponent\cite{Chalker-Coddington,Huckestein-95,Mieck-90}.

Were it not for localization, 
propagation of a wavepacket in a random
potential would be diffusive,
\begin{equation}
  \langle \Delta R^2(t)\rangle\sim Dt.
  \label{eq:diffusive-spread}
\end{equation}
Localization corrections are least important near the critical energy,
but even there they modify the form of a wavepacket at very large
times\cite{Chalker-Daniell-88}.
However, for not too large times the
r.m.s.\ displacement, which is primarily determined by small momenta, retains the
diffusive form.  This can be used to find the
conductivity at small frequencies.  

To this end, let us rewrite
Eq.~(\ref{eq:kubo-guiding})
\begin{equation}
  \label{eq:kubo-rms}
    \tilde\sigma(\omega)=
  -\frac{2\omega^2}{l^2\gamma}\lim_{\delta\to +0}\re\int_{0}^{\infty}dt\,
  e^{i\omega t-\delta t}\,{\rm Tr}_0\,\overline{
    \Delta R^2(t)
    }, 
\end{equation}
in terms of the squared displacement $\Delta R^2(t)\equiv \left[{\bf
R}(t)-{\bf R}(0)\right]^2$, where ${\bf R}(t)\equiv e^{iHt} {\bf
R}\,e^{-iHt}$ is the Heisenberg operator of the guiding center.  For
an eigenstate $|n\rangle$ of the Hamiltonian~(\ref{eq:hamiltonian})
randomly chosen not too far from the critical energy, $\langle n|
\Delta R^2(t)|n \rangle$ has the diffusive
form~(\ref{eq:diffusive-spread}) at small enough $t$, but it
eventually saturates at the distance of the order of the localization
length $\xi_{\varepsilon_n}$.
Replacing the trace by the integral over energy weighted with the
(non-critical) density of states, we obtain the overall long-time
($\gamma t\gg1$) r.m.s.\ displacement
\begin{eqnarray}
  \label{eq:average-displacement}
  {\rm Tr}_0\overline{\Delta R^2(t)}& \sim&\hbar\gamma
  \!\int\! d\varepsilon
  \,\rho(\hbar\gamma\varepsilon)\min(Dt,\xi^2_\varepsilon)
  \nonumber\\
  &\propto&
  l^2\,(\gamma t)^{1-1/(2\nu)}. 
\end{eqnarray}
This average is determined by the states with energies $|\varepsilon|
\lesssim (\gamma t)^{-1/2\nu}$; the integral rapidly converges outside
this region.   

With asymptote (\ref{eq:average-displacement}), time integration in
Eq.~(\ref{eq:kubo-rms}) gives:
\begin{equation}
  \label{eq:scaling-sigma-large-temperature}
  \tilde\sigma(\omega)=C\,(\omega/\gamma)^\mu, 
  \quad 
  \mu\equiv (2\nu)^{-1}.
\end{equation}

The same result can be obtained from the scaling
form\cite{Wang-Fisher-Girvin-Chalker,Sondhi-unpublished} of the
zero-temperature conductivity of the non-interacting system at a given
chemical potential, which can be written as
\begin{equation}
  \label{eq:scaling-sigma-zero-temperature}
  \sigma_{xx}(\varepsilon,\omega)={e^2\over \hbar}\,{\cal
    G}_0\left({\omega \xi_{\varepsilon}^{2}\over 
      \gamma l^2}\right),
\end{equation}
where the dimensionless scaling function ${\cal G}_0(X)$ rapidly
vanishes for $X\to 0$ 
and approaches a constant value for large $X$.  Indeed, the
conductivity 
for $\beta\omega\ll1$
can be written as a convolution of the scaling
function~(\ref{eq:scaling-sigma-zero-temperature}) with the derivative
of the Fermi distribution function 
\begin{equation}
  \tilde\sigma(\omega)={8 \,k_BT\over ne^2\,l^2\gamma}
  \int d\varepsilon \,\Bigl(-{dn_F\over
    d\varepsilon}\Bigr)
    \sigma_{xx}(\varepsilon,\omega)
  \label{eq:conductivity-answer}
\end{equation}
[cf.\ Eq.~(\ref{eq:einstein-relation})].
For $k_BT\gg \hbar\gamma$, all energies within the stripe of width
$\delta\varepsilon\sim(\omega/\gamma)^\mu$ contribute equally, and in
the limit $\omega\to0$ we
obtain Eq.~(\ref{eq:scaling-sigma-large-temperature}) with the coefficient
\begin{equation}
  C= {16\pi \mu}
  \int_{-\infty}^\infty {dX\over |X|^{1+\mu}} \,\tilde{\cal G}_0(X).
  \label{eq:conductivity-ans-coeff}  
\end{equation}
Here we have assumed that $\varepsilon^{\nu}\xi_{\varepsilon} \to
$~const for $\varepsilon\to 0$, and $\tilde {\cal G}_0(X)\equiv
\lim_{\varepsilon\to0}{\cal G}_0(X\,\varepsilon^{2\nu}
\xi_{\varepsilon}^{2}/ l^2)$.  The integration converges both at zero
and infinity.



\subsection{Spectral moments}
\label{subsec:moments}

Since the single-particle conductivity goes to zero both for
$\omega\to 0$ and for $\omega\gg \gamma$, its frequency dependence
displays a peak, with a maximum at a nonzero frequency $\omega \sim
\gamma$. Such a peak is of central interest from the point of view of
experiment, it does not occur in the SCBA. This peak was previously
found and briefly discussed in our communication \cite{Kuehnel-00}.
Here we present the results and provide some details of the full
calculation of the low-frequency conductivity based on the method of
spectral moments (MOM). The advantageous feature of this method is
that, instead of solving the full time-dependent problem of the
electron motion in a random field (\ref{eq:guiding-velocity}), one has
to evaluate only equal-time correlation functions.

We first calculate the spectral moments of the reduced conductivity
$\tilde\sigma(\omega)$ [Eq.~(\ref{eq:einstein-relation})]. They are
defined as 
\begin{equation}
  \label{MOM}
  M_k={1\over 2\pi\gamma}\int_{-\infty}^{\infty} d\omega
  \,(\omega/\gamma)^k\tilde\sigma(\omega). 
\end{equation}
For $\omega, \gamma\ll k_BT/\hbar$, the states within the broadened
Landau level are equally populated, and the conductivity
is an even function of frequency, $\tilde\sigma(\omega) =
\tilde\sigma(-\omega)$. Therefore all odd moments vanish, $M_{2k+1}
=0$. For even moments, we use the Hamiltonian~(\ref{eq:hamiltonian})
to obtain from Eqns.~(\ref{eq:kubo-formula2}), (\ref{MOM}),
\begin{eqnarray}
  \label{general_MOM}
  M_{2k}=&&-2l^2
  \sum
  ({\bf q}_1\,{\bf q}_{2k+2})\,\overline{\tilde V_{{\bf q}_1}\ldots 
  \tilde V_{{\bf q}_{2k+2}}}\\
  &&\times \left[\left[\ldots\left[e^{i{\bf q}_1{\bf R}},\,
              e^{i{\bf q}_2{\bf R}}\right],\ldots\right],\,
      e^{i{\bf q}_{2k+1}{\bf R}}\right]\,e^{i{\bf q}_{2k+2}{\bf
  R}}.\nonumber 
\end{eqnarray}
The summation is performed over all ${\bf q}_1,\ldots,{\bf q}_{2k+2}$.
The commutators (\ref{general_MOM}) can be evaluated recursively using
\begin{equation}
  \label{commutator}
  \bigl[e^{i{\bf qR}},\,e^{i{\bf q}'{\bf R}}\bigr]=
      2i\sin\Bigl({1\over2}l^2{\bf 
      q}\wedge{\bf q}'\Bigr)\, e^{i({\bf q}+{\bf q}')\,{\bf R}}.
\end{equation}

For Gaussian random potential, the disorder average in
Eq.~(\ref{general_MOM}) can be computed by Wick's theorem.  From
Eqns.~(\ref{eq:effective-potential-lll})--(\ref{potential})
$$
\langle \tilde V_{\bf q}\tilde V_{{\bf
q}'}\rangle = (\pi l^2/2S)\,\exp(-l^2q^2/2)\,\delta_{{\bf q}+{\bf q}',{\bf 0}},
$$
where $S$ is the area of the system.  Then,
\begin{eqnarray}
M_{2k} &=&
 \label{eq:mom_sinus}
 -\pi\left(-{l^2 \over 2\pi} \right)^{k+2}\sum_{
             {\cal C}( \left\{ {\bf q} \right\} )  }
 \int d{\bf q}_1 \cdots d{\bf q}_{2k+2}\, {\cal C}( \left\{ {\bf q}
 \right\} )
 \nonumber \\
 & & \times ({\bf q}_1 {\bf q}_{2k+2})
 \exp\Biglb( -{l^2\over 4} (q_1^2+\ldots+q_{2k+2}^2) \Bigrb) \nonumber \\
 & & \times \sin\Bigl({l^2 \over 2}\,{\bf q}_1 \wedge {\bf q}_2\Bigr)
 \,\sin\Biglb({l^2 \over 2}\,({\bf q}_1+{\bf q}_2)\wedge {\bf q}_3\Bigrb)
  \ldots\nonumber \\ 
 & &   \times
 \sin\Biglb({l^2 \over 2}\,({\bf q}_1+{\bf q}_2+\ldots+{\bf q}_{2k})
 \wedge{\bf q}_{2k+1}\Bigrb)
\end{eqnarray}
where the sum is taken over all $(2k+1)!!$ ways to choose pairs out of
the set of $2k+2$ variables, and 
$${\cal C}(\left\{ {\bf q} \right\} )
\equiv \delta({\bf q}_{i_1} + {\bf q}_{j_1}) \ldots \delta({\bf
  q}_{i_{k+1}} + {\bf q}_{j_{k+1}})
$$
is the corresponding {\em contraction\/} function. 
\begin{figure}[htb]
\strut\hfill
\epsfxsize=0.7\columnwidth%
\epsfbox{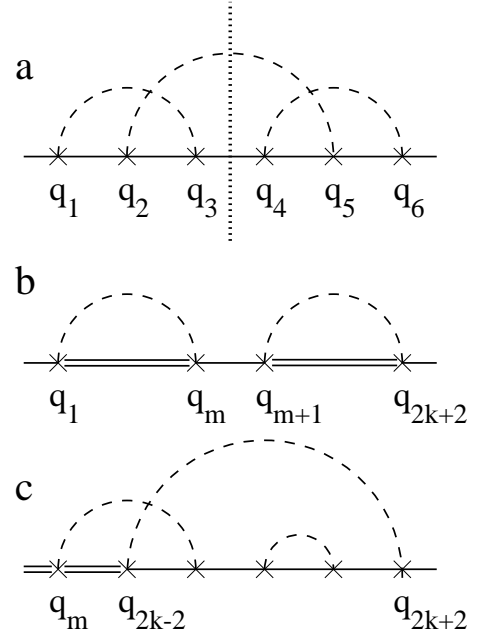}
\hfill\strut
\caption{Examples of diagrams. Dashed lines indicate which variables
are paired, double lines represent an arbitrary internal
structure. (a) A symmetric diagram.  Diagrams whose contribution is
equal to zero: (b) {\em disconnected\/} diagrams, and (c) diagrams
which vanish because the associated function is odd in ${\bf
q}_{2k+2}$.}
\label{fig:zero_diagram}
\end{figure}

To classify different terms in the sum~(\ref{eq:mom_sinus}), it is
convenient to depict the contraction procedure graphically as
illustrated in Fig.~\ref{fig:zero_diagram}.  (These diagrams merely
represent the contraction function $\cal C$ and should not be confused
with Feynman diagrams for the Green's functions.)  First, we
note that the sequence of ${\bf q}_i$ paired in a diagram may be
reversed, ${\bf q}_i\to {\bf q}_{2k+3-i}$, without changing the
overall value of the diagram.  The diagrams obtained by such a
reversal are equivalent, which reduces the computation cost by a
factor of two [this reduction does not occur, however, for symmetric
diagrams shown in Fig.~\ref{fig:zero_diagram}a]. Additional
simplification comes from the fact that disconnected diagrams
(Fig.~\ref{fig:zero_diagram}b) and the diagrams with the structure
shown in Fig.~\ref{fig:zero_diagram}c are equal to zero. The number of
diagrams of different sorts is given in Ref.~\CITE{Kuehnel-thesis};
for example, for the 14th moment there are 2027035 diagrams, out of
which 5937 are symmetric, 318631 are disconnected; the contribution of
the diagrams with ${\bf q}_1=-{\bf q}_2$ is $\approx 72.559$, whereas
the contribution of all other diagrams is $\approx -8.809$.

Despite the reductions, the number of terms to be calculated remains
very large for large $k$.  Moreover, each term in
Eq.~(\ref{eq:mom_sinus}) is a sum of $2^{2k}$ Gaussian integrals.
Each integral can be calculated algebraically but at a high computational
cost.  To accelerate the calculation, we have devised an efficient
numeric classification scheme, which sorts diagrams inexpensively into
{\em bins\/} according to their approximate values calculated with
double precision.  A representative diagram is evaluated algebraically
for each bin. Finally, the diagrams are summed up with proper
multiplicity, giving {\it exact} numerical values of the moments.  The
procedure is outlined in Appendix~\ref{app:classification}.
Calculating algebraically only non-equivalent Gaussian integrals
reduces computational time tremendously. This allowed us to evaluate the
moments up to $M_{14}$.  For $k=0,1,\ldots,7$ we obtain

\wid

\begin{eqnarray}
\label{m_values}
&&\!\!\!\!\!\!M_{2k}=1; {3\over 8}; {443 \over 1152}; {25003\over 38400};
{13608949709 \over 8941363200 };
{298681273551508807 \over 66698308912435200};
{566602308094143977186611746328323669809 \over
36033364452669289726755567308636160000}; \nonumber \\
&&\qquad\quad
{2589008911677049308284617052653287524724669331093372792412270459939701 \over
    40611974008223423608381355617240666314144290787406293503186042880000 }
\end{eqnarray}

\nar

and the corresponding approximate values,
\[
M_{2k}\approx 1;0.375; 0.385; 0.651; 1.522; 4.478; 15.72; 63.75\;.
\]

\subsection{Reconstruction of frequency-dependence}

\begin{figure}[htb]
\begin{center}
\strut\hfill%
\epsfxsize=0.7\columnwidth
\epsfbox{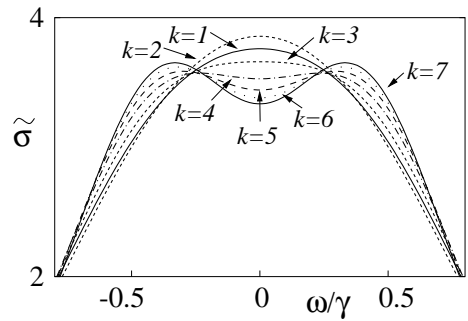}\hfill\strut
\end{center}
\caption{Approximating $\tilde{\sigma}(\omega)$ with Hermite
polynomials. With the increasing number of moments $M_{2k}$, a minimum of
the conductivity is consistently developing at $\omega=0$, and the
expansion does not show fast convergence for small $\omega/\gamma$.}
\label{fig:bad_sigma}
\end{figure}

Since the conductivity is asymptotically Gaussian, one is tempted to
restore $\tilde\sigma(\omega)$ from the moments $M_n$ in a standard
way, writing an expansion in Hermite polynomials $\tilde\sigma(\gamma
x)=\sum_n B_n\,H_n(\sqrt 2 x)\exp(-2 x^2)$.  The coefficients $B_n$ can
be expressed recursively in terms of the moments $M_{k}$, $k\le n$.
However, for the moments~(\ref{m_values}), this expansion does not
converge rapidly, see Fig.~\ref{fig:bad_sigma}.  This is consistent
with {\em nonanalyticity\/} of the conductivity at $\omega= 0$.

Given the exponent $\mu$ in
Eq.~(\ref{eq:scaling-sigma-large-temperature}), a much more rapidly
convergent expansion can be constructed in terms of a different set of
orthogonal polynomials.  Specifically, with
Eqns.~(\ref{eq:large-omega}),
(\ref{eq:scaling-sigma-large-temperature}), we write the conductivity
at the lowest Landau level as
\begin{equation}
  \label{G-function}
  \tilde\sigma(\omega)=x^{\mu}G(x)
  \exp(-2x^2),\;x\equiv |\omega|/\gamma.  
\end{equation}
The function $G(x)$ ($x \geq 0$) can be expanded in Laguerre
polynomials $L_n^{(\mu-1)/2}(2x^2)$, which are orthogonal for the
weighting factor in Eq.~(\ref{G-function}). It is important that the
expansion coefficients can also be recursively restored from the
moments $M_{2k},\, k\leq n $.

\begin{figure}[htb]
\begin{center}
\epsfxsize=\columnwidth
\epsfbox{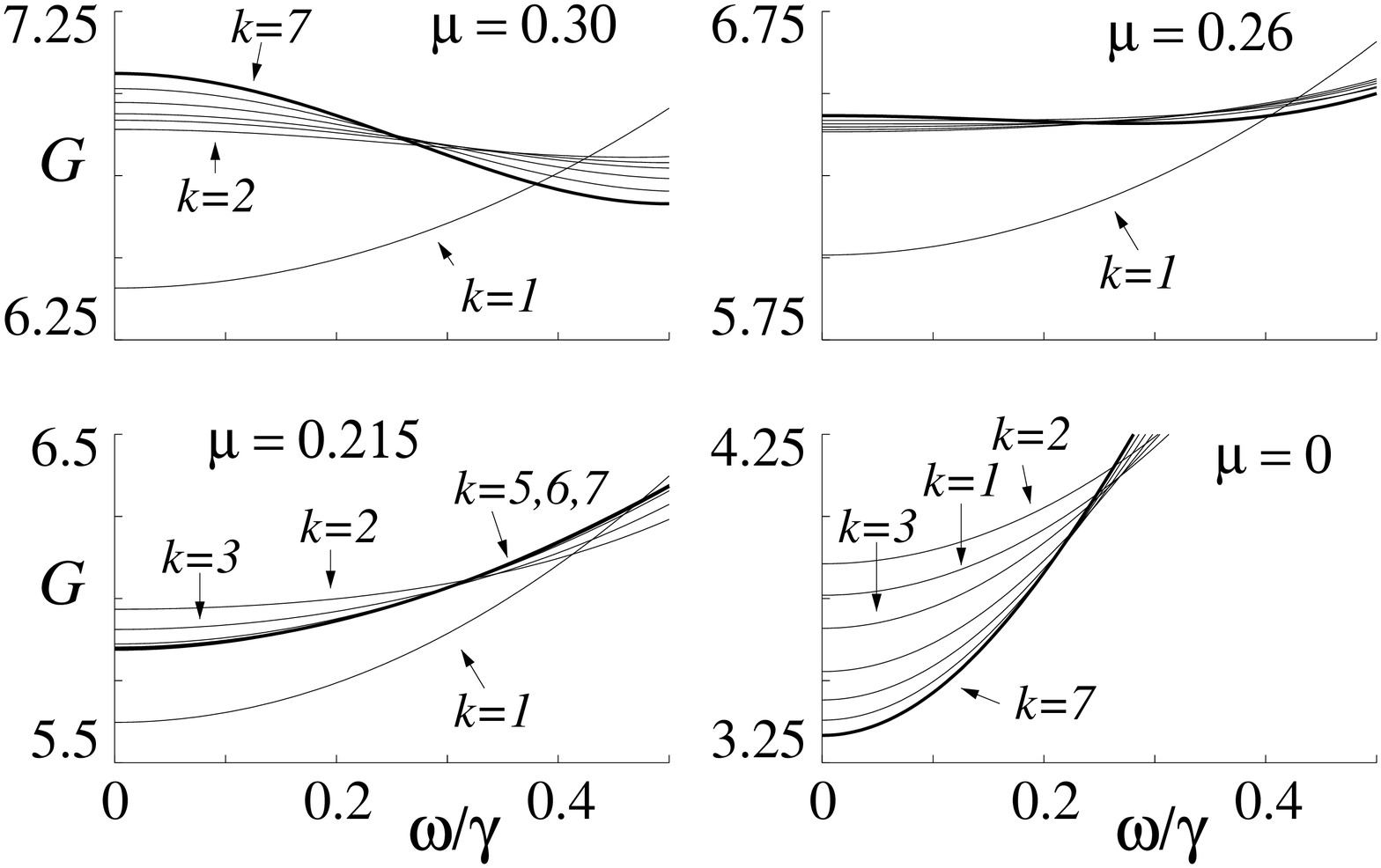}
\end{center}
\caption{The expansion of the prefactor $G$ (\protect\ref{G-function})
in Laguerre polynomials $L_n^{(\mu-1)/2}(2x^2)$ for different values
of the exponent $\mu$, depending on the total number of moments
$M_{2k}$.  The expansion converges rapidly for $\mu$ between 0.19 and
0.28 and much slower outside this interval.}
\label{fig:prefactor}
\end{figure}

For the presently accepted value of the localization exponent
$\nu\approx 2.33$, the value of the conductivity exponent $\mu$ is
0.215. The expansion for $G$ converges rapidly for $\mu$ between
$0.19$ and $0.28$, whereas outside this region the convergence
deteriorates, as illustrated in Fig.~\ref{fig:prefactor}. This could
be considered as an indirect indication of the consistency of our approach.

The resulting conductivity calculated with $\mu=0.215$ is shown in
Fig.~\ref{fig:conduct}.  The estimated deviation from the obtained
curve due to the finite number of moments and also to the uncertainty
in the value of $\mu$ (its effect is discussed in
Ref.~\CITE{Kuehnel-thesis}) is smaller then the width of the line.

\section{Single-electron cyclotron resonance}
\label{sec:cyclotron}

Resonant single-electron absorption near cyclotron frequency
($\omega\approx \omega_c$) is determined by the correlation function
of the Landau level raising and lowering operators $p_\pm$ defined in
Appendix~\ref{app:guiding},
\begin{equation}
  \label{eq:cyclotron-original}
  \sigma_c(\omega) = {ne^2
  \over 2 m} \int_{-\infty}^\infty dt\; e^{i\omega t}
  \; \langle p_{-}(t) p_{+}(0) \rangle, 
\end{equation}
where we used Eq.~(\ref{eq:kubo-full}) and assumed that
$\exp(\hbar\omega_c/k_BT)\gg 1$, in which case only the lowest Landau
level is occupied for small densities. If the disorder is weak, $\gamma\ll
\omega_c$, it only weakly mixes different Landau levels, the primary
effect being lifting the degeneracy of each level.  Then, the problem
of optically-induced transitions between different Landau levels
resembles that of transitions between degenerate electronic terms of
impurities in solids in the presence of the electron-phonon coupling
which gives rise to Jahn-Teller effect\cite{Stoneham-75}. One of the
effective methods of the theory of absorption spectra of Jahn-Teller
centers is the method of moments.

The MOM formalism can be transferred to the case of
inter-Landau-level transitions.  For
$\hbar\gamma\ll k_BT$, quenched disorder can be described in the same
way as scattering by thermally excited phonons.  The major difference
is infinite level degeneracy.

In the neglect of disorder-induced scattering between Landau levels,
one should keep only the part $H_d$ of the disorder potential $V({\bf
r})$, which is diagonal in the Landau level representation,
\begin{equation}
   \label{Hi_represent}
   H_d =\sum_N H_d^{(N)}\,P_N
   = \hbar\gamma \sum_{\bf q} \tilde V_q e^{i{\bf q R}}
   \sum_N L_N\Bigl({q^2\,l^2\over2}\Bigr)\,P_N, 
\end{equation}
where $\tilde V_q$ is defined by
Eq.~(\ref{eq:effective-potential-lll}) and $P_N=P_N^2$ is the operator
of projection to the $N$\,th Landau level, as in
Eq.~(\ref{eq:projection-operation}).  With the Hamiltonian
(\ref{Hi_represent}), 
oscillations at the cyclotron frequency can be singled out in
Eq.~(\ref{eq:cyclotron-original}),
\begin{equation}
  p_{\pm}(t) = e^{\pm i\omega_c t} e^{i H_d t/\hbar} p_{\pm} 
e^{-i H_d t/\hbar}.
\end{equation}
Then, from Eq.~(\ref{eq:cyclotron-original}), we can write
\begin{equation}
  \sigma_c(\omega) =
  {ne^2\over
    2m\gamma}\,\tilde\sigma_c(\omega),
  \label{eq:sigma-cyclotron-coeff}
\end{equation}
where 
\begin{equation}
  \label{red_sigma_cyc}
  \tilde \sigma_c(\omega) = \gamma\! \int_{-\infty}^\infty \!dt\,
  e^{i\Delta\omega\,t}\, {\rm Tr}_0\overline{\left[
    e^{i H_d t/\hbar} p_{-} e^{-i H_d t/\hbar}
  p_{+}\right]}
\end{equation}
is the reduced conductivity, and $\Delta\omega\equiv \omega-\omega_c$ is
the frequency detuning, $|\Delta\omega|\ll \omega_c$.

The major difference of Eq.~(\ref{red_sigma_cyc}) from its counterpart
(\ref{eq:kubo-guiding}) for the low-frequency conductivity is that the
Hamiltonians for direct and inverse time propagation (corresponding to
the factors $e^{\pm i H_d t}$) are now different, which is again
familiar from the theory of impurity absorption spectra.  The reduced
cyclotron conductivity can be conveniently written in a form
conventional for this theory by introducing the ``perturbation''
Hamiltonian
\begin{equation}
  \delta H_d\equiv H_d^{(1)} -H_d^{(0)}
  =-\hbar\gamma \sum_{\bf q} {q^2 l^2\over 2}\,
    \tilde V_{\bf q} \,e^{i{\bf q R}}.
    \label{eq:perturbation-cyclotron}
\end{equation}
In the interaction representation, $\tilde\sigma_c$ can be then simply
expressed in terms of a time-ordered exponential,
\begin{equation}
  \label{red_sigma_cyc2}
  \tilde \sigma_c(\omega) = \gamma \!\int_{-\infty}^\infty \!dt\,
  e^{i\Delta\omega\, t}\,
 {\rm Tr}_0\biggl[\overline{ {\rm T}_\tau
  \exp\Biglb(-{i\over \hbar}\!\int\nolimits_{0}^t \! d\tau\,\delta 
  H_d(\tau)
 \Bigrb)} \biggr]. 
\end{equation}
Here, time dependence of the operator $\delta H_d$,
\begin{equation}
 \delta H_d(\tau) \equiv  e^{i H \tau/\hbar} \,\delta H_d \, 
e^{-i H \tau/\hbar}, \quad 
H\equiv H_d^{(0)},
\end{equation}
is generated by the disorder Hamiltonian projected on the LLL, which is
given by Eq.~(\ref{eq:hamiltonian}) of the previous section. 

We can now define the spectral moments of the cyclotron peak as
\begin{equation}
  \label{cyclotron_general_MOM}
  M_k^c={1\over 2\pi\gamma}\int_{-\infty}^{\infty} d\omega
  \,\left( \omega - \omega_c \over \gamma \right)^k\tilde\sigma_c(\omega).
\end{equation}
Using Eq.~(\ref{red_sigma_cyc2}) we write
\begin{equation}
  \label{mom_cyc2}
  M_k^c = 
 {\rm Tr}_0\overline{\left({i\over\gamma}\,{d\over dt}\right)^k {\rm T}_{\tau}
 \exp\Biglb(-{i\over \hbar}\int\nolimits_{0}^t 
 \delta H_d(\tau) \; 
  d\tau\Bigrb)}\Biggr|_{t=0}. 
\end{equation}
We note that, similar to the case of the peak of low-frequency
conductivity discussed in the previous section, we are calculating
here the moments of the cyclotron peak only, whereas the small
($\propto \gamma/\omega_c$) background from the correlators neglected
in obtaining Eq.~(\ref{eq:cyclotron-original}) is projected away, as
are also the peaks of $\sigma_{xx}(\omega)$ at $\omega\approx
n\omega_c$ with $n\neq 1$.

\subsection{Tails of the cyclotron resonance peak}

As in the previous section, let us first discuss
the asymptotic form of the cyclotron peak comparatively far from
resonance, $|\Delta\omega|\gg\gamma$ (yet $|\Delta\omega|\ll
\omega_c$).  If we introduce the exact eigenstates of the
Hamiltonian~(\ref{Hi_represent}) for the lowest $|0,m\rangle$ and the
first excited $|1,m\rangle$ Landau levels, with energies $E_m^{(0)}$
and $E_m^{(1)}$, respectively, the expression~(\ref{red_sigma_cyc})
for the reduced conductivity can be written in the form
\begin{equation}
  \label{cyclotron_absorption}
  \tilde\sigma_c(\omega) = {2\pi\hbar\gamma} \sum_{m,n}\overline{
  \delta(E_m^{(1)} - E_n^{(0)} - \hbar\omega) | \langle 1, m| p_{+} |0,
  n\rangle |^2}. 
\end{equation}
As for the low-frequency conductivity considered in the previous
section, the conductivity tail is determined by large optimal
fluctuations of the disorder potential.

The problem of the optimal potential for cyclotron
resonance was previously considered by Ioffe and
Larkin\cite{Ioffe-81}.  They used an ansatz of a rotationally-symmetric
optimal potential
\begin{equation}
   \label{Larkin_solution}
   V_{{\rm opt}}^{{\rm IL}} =
   2\pi V_0 |\Phi_0|^2 + 2\pi V_1 |\Phi_1|^2,
\end{equation}
where
$  \Phi_0 = \psi_{0,0}({\bf r}),\;    \Phi_1 = \psi_{1,-1}({\bf r})
$
are the functions of the lowest and first excited Landau levels
centered at the {\it same} origin, with magnetic quantum numbers 0 and $-1$,
respectively.  This resulted in the asymptotic form of the cyclotron
resonance absorption peak $\sigma_c \propto \exp(-8\,
\Delta\omega^2/\gamma^2)$, for the range $\hbar\gamma\ll k_BT$.

We argue that the transition probability between the states with
energy separation $E_m^{(1)}-E_n^{(0)} = \Delta\omega + \omega_c$ is
exponentially increased if the cyclotron orbit centers of these states
are permitted to shift with respect to each other.  This happens
despite the associated decrease of the overlap integral. 

The calculation of the tails of the cyclotron resonance absorption
peak is very similar to that in Sec.~\ref{sec:low-tails}.  We begin by
writing the averaging in terms of a functional integral
(\ref{eq:path-integral}), with the energy conservation taken into
account using a Lagrange multiplier [as in
Eq.~(\ref{eq:variational-sigma-two}) but with {\em different\/}
Hamiltonians for $E_t$ and $E_b$]. If we neglect the dependence of the
transition matrix element on $V({\bf r})$, then for the optimal
potential we obtain an equation
similar to Eq.~(\ref{Larkin_solution}). However, in contrast to
Ref.~\CITE{Ioffe-81}, we permit the centers of the wave functions
$\Phi_0$ and $\Phi_1$ to be shifted with respect to each other. 

A remarkable feature of this simplified variational problem is that,
in the neglect of overlapping of the displaced wave functions, the
{\it same} value of the variational functional [except for the overlap
term] is obtained for the trial wave functions of the first Landau level
with the magnetic quantum numbers -1 or 0, \ie, $\psi_{1,-1}$
or $\psi_{1,0}$, or for any their linear combination.

For a displacement $R$ between the centers of the hump and well of the
optimal potential, the transition matrix element is $|\langle \psi_1 |
p_{+} | \psi_0 \rangle| \sim \exp(-R^2/4l^2)$. The optimal distance $
R^2 \approx 4l^2 \ln [(\omega-\omega_c)^2/\gamma^2]$ is found by maximizing
the expression with the matrix element present.  As in the case of the
low-frequency conductivity, this distance increases as the frequency
is tuned away from resonance.

The variational result for the conductivity tail is
\begin{equation}
  \label{cyclotron_asymptotic}
  \tilde \sigma_c(\omega) \propto \exp\Biglb(-{8\over
  3\gamma^2}(\omega-\omega_c)^2
  \Bigrb). 
\end{equation}
This tail is much broader, 
with the exponent reduced by a factor of 3, compared to the result
of Ref.~\CITE{Ioffe-81}.

\subsection{The center of the cyclotron absorption peak}

Generally, we do not expect $\tilde\sigma_c(\omega)$ to display a
non-analytic dip at the center of the cyclotron absorption peak.
Indeed, the power-law
singularity~(\ref{eq:scaling-sigma-large-temperature}) of the
low-frequency conductivity can be associated with quantum interference
which leads to eventual localization of all states except for one (or
maybe a few) at the band center.  The
expression~(\ref{red_sigma_cyc2}) for the cyclotron resonance
absorption has a structure which differs from that for the
low-frequency conductivity. In particular, it contains an extra phase
factor from the Hamiltonian $\delta H_d$
(\ref{eq:perturbation-cyclotron}) which represents the difference in
the random potential experienced by an electron at the two Landau
levels.  This phase factor should give rise to an exponential damping,
and related suppression of the interference effects at long times.
Consequently, the conductivity is expected to be smooth near
$\omega_c$.  

Another way to see this is based on the following arguments.  The
suppression of the low-frequency conductivity for $\omega\to 0$ may be
attributed to level repulsion between overlapping localized states.
This repulsion is comparatively small for states of relatively large
radii, with energies close to the band center. Indeed, only such
states contribute to the low-frequency conductivity, as we saw in
Sec.~\ref{sec:small-omega}. On the other hand, resonant cyclotron
absorption is due to transitions between {\em different\/} Landau
levels. Although the central part of the absorption peak is formed by
transitions between strongly overlapping states, the involved states
are eigenstates of {\em different\/} Hamiltonians, with random parts
$H_d^{(0)}$ and $H_d^{(1)}$. Their wave functions have different
spatial structures and their energies are essentially uncorrelated,
except for states deep in the tails of the Landau levels.
Consequently, we expect no suppression of transitions at frequencies
close to $\omega_c$. This argument is in agreement with the results of
the method of moments presented in the next subsection and
Fig.~\ref{fig:cyc_conduct}.

\subsection{Spectral moments of the cyclotron peak}
We will now calculate the spectral moments~(\ref{mom_cyc2}).  Because
all states of the lowest Landau level are equally populated, the
reduced conductivity~(\ref{red_sigma_cyc2}) is symmetric with respect
to $\omega_c$, \ie, $\tilde\sigma_c(\omega_c + \Delta\omega) =
\tilde\sigma_c(\omega_c -\Delta\omega)$. Therefore all odd moments
vanish, $M^c_{2k+1} =0$.  The structure of the expression for even
moments which follows from Eq.~(\ref{mom_cyc2}) is similar to that of
Eq.~(\ref{uncontracted}).  The main difference is that the prefactor
now is a complicated polynomial, a combination of products of terms
which are linear in the squared wave numbers $q_k^2$.

This strongly complicated the numerical procedure. In particular, we
failed to find any symmetries to reduce the computational overhead,
and graphical representations were of little help. We were also unable to
categorize different terms as described in
Appendix~\ref{app:classification} for the low-frequency conductivity.
Instead, we developed the computer algebra package {\tt
GaussInt}\cite{Mathematica} for Mathematica, capable of handling the
integration of high dimensional Gaussian integrals in a manageable
time frame, and used the brute-force approach calculating all terms
analytically.  For $k=0,1,\ldots,5$ we obtained:
\begin{eqnarray}
\label{mcyc_values}
M^c_{2k}=&&1; {1 \over 2}; {37 \over 64}; {52043 \over 55296};
{4750893001499 \over 2488320000000}; \nonumber \\
&&\hspace{-0.8cm}
{29694054188353275207831950716496054687 \over
  6480696333914117611721116876800000000}, 
\end{eqnarray}
and the corresponding approximate values
\[
M^c_{2k}\approx 1.000; 0.500; 0.578; 0.941; 1.909; 4.582.
\]
The values of the moments with $k=0,1,2$ were also independently confirmed
analytically.

\subsection{Reconstruction of frequency-dependence}

\begin{figure}[htb]
\begin{center}\strut\hfill
\epsfxsize=0.9\columnwidth
\epsfbox{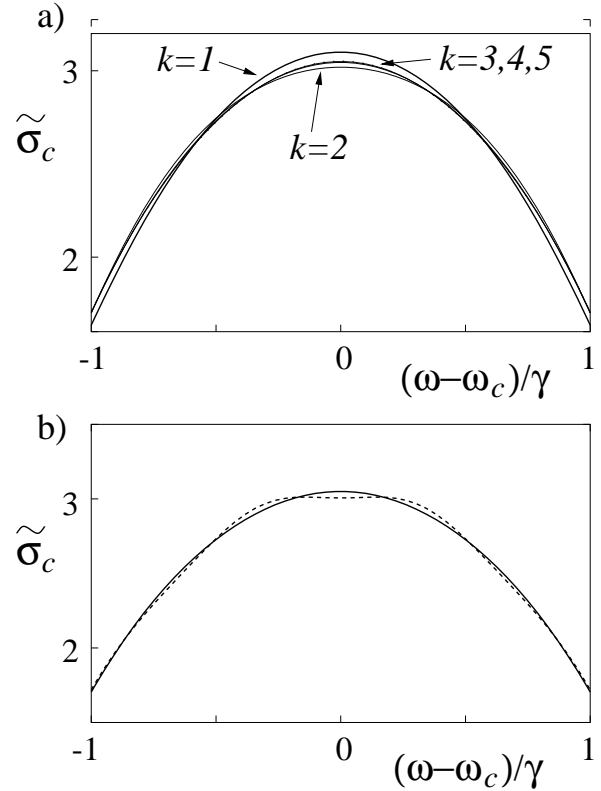}\hfill\strut
\end{center}
\caption{Approximating $\tilde{\sigma}_c$ with continued fractions.
(a) The convergence with increasing number of moments $M^c_{2k}$ is
extremely fast: the curves lie on top of each other already for
$k=3,4,5$.  (b) A comparison between the continued-fraction (solid
line) and the Hermite polynomial approximation (dashed line) for
$k=5$.  Convergence is much faster with continued fractions.}
\label{fig:sigma_cyc_cfraction}
\end{figure}

As a first step, we  reconstructed $\tilde\sigma_c(\omega)$
using a standard expansion in Hermite polynomials,
$$
\tilde\sigma_c(\gamma x)=\sum_n
B_n\,H_n(\sqrt{8/3}\, x)\exp(-8x^2/3).  
$$
The coefficients $B_n$ were recursively expressed from the calculated
moments~(\ref{mcyc_values}).  We discovered that although the
convergence is fast far from the center of the peak, it is noticeably
slower close to the center [we emphasize, however, that we reached
convergence, in contrast to the similar expansion for the
low-frequency conductivity in Fig.~\ref{fig:bad_sigma}, where the
convergence was not reached for 14 moments]. Within this approach, the
number of calculated moments is apparently insufficient for restoring
the entire function $\tilde\sigma_c$ with desired accuracy.  The
corresponding result is shown with dashed line in
Fig.~\ref{fig:sigma_cyc_cfraction}).

Much faster convergence was achieved when $\tilde\sigma_c(\omega)$ was
restored using a continued fraction expansion. We applied an algorithm
similar to that used to reconstruct the LLL density of states from its
frequency moments for an arbitrary correlated random
potential\cite{Bohm-97}.  The steps involved in this process are
summarized in Appendix~\ref{app:continued-fraction}.  As one can see
from Fig.~\ref{fig:sigma_cyc_cfraction}, the convergence is very fast.
The resulting shape of the cyclotron absorption peak is shown in
Fig.~\ref{fig:cyc_conduct}.  We believe that the deviation from the
exact value is within the width of the curve.

\section{conclusion}
In conclusion, we have analyzed the low-frequency single-electron
magnetoconductivity and cyclotron resonance absorption of a
nondegenerate 2D electron system in a quantizing magnetic field. We
considered the experimentally important parameter range where the
width of the Landau levels is less than temperature, so that all
states within the lowest Landau level are equally populated. In this
range, by combining the ideas of the scaling theory of the IQHE, the
method of optimal fluctuation, and the method of spectral moments, we
obtained highly accurate numerical results throughout the frequency
domain where the conductivity displays peaks.  

We found that, in contrast to the prediction of the SCBA or other
mean-field theories \cite{Saitoh-87}, the low-frequency conductivity
displays a peak at a {\it nonzero} frequency, as shown in
Fig.~\ref{fig:conduct}. For short-range disorder, the position of the
peak is given by
\begin{equation}
\label{eq:peak-position}
\omega_{\rm m}\approx 0.26\gamma.
\end{equation}
For $\omega\to 0$, the single-electron conductivity displays a {\it
universal} power-law dispersion $\sigma_{xx}\propto \omega^{\mu}$,
which is related to the scaling behavior of the localization length as
a function of the distance in energy from the center of the
disorder-broadened Landau level. On the other hand, the peak of the
cyclotron resonance does not display such singular behavior and is not
shifted away from $\omega_c$, as seen from Fig.~\ref{fig:cyc_conduct}.
Both peaks have Gaussian tails, with different exponents
[see Eqns.~(\ref{eq:large-omega}), (\ref{cyclotron_asymptotic})].

\begin{figure}
  \begin{center} 
  \epsfxsize=\columnwidth \epsfbox{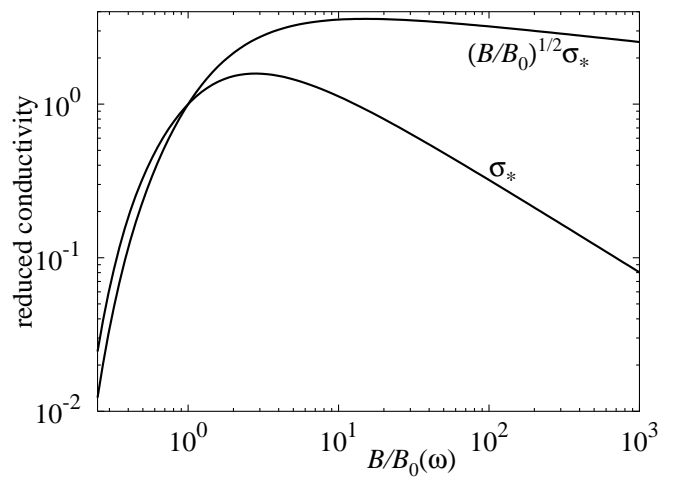}
  \end{center} 
\caption{Reduced ac magnetoconductivity $\sigma_*$
(\protect\ref{eq:scaled-B}) at a nonzero frequency $\omega$ as a
function of the reduced magnetic field $B/B_0(\omega)\propto
B\,\omega^{-2}$.  In order to demonstrate the anomalous single-electron
behavior,  $\sigma_*$ is also plotted with an
extra factor $({B/B_0})^{1/2}$.  For large $B$, the
single-electron conductivity displays scaling behavior,
${B}^{1/2}\sigma_* \propto B^{-\mu/2}$.  }
\label{B-dependence}
\end{figure}

Experimentally, it is more feasible to investigate the
magnetoconductivity at a given nonzero frequency $\omega$ as a
function of the external magnetic field $B$.  The corresponding
representation of our results is given in Fig.~\ref{B-dependence} for
the scaled conductivity $\sigma_*(B;\omega)$,
\begin{equation}
\label{eq:scaled-B}
\sigma_*\equiv \sigma_*(B;\omega)= \left[{B_0(\omega)\over B}\right]^{1/2}
{\tilde\sigma(\omega)\over \tilde\sigma(\gamma)},
\end{equation}
where the scaling factor $\tilde\sigma(\gamma)\approx 1.08$, and the
scaling field $B_0(\omega)$ is defined by the equation
$\gamma(B_0)=\omega$. The magnetoconductivity $\sigma_{xx}$ is related
to $\sigma_*(B;\omega)$ by a factor which is independent of $B$ (but depends on
$\omega$),
\begin{equation}
  \label{eq:sigma*}
  \sigma_{xx}(\omega)= {\tilde\sigma(\gamma)\over 4\pi}\,
  {\hbar\over k_B T}\, {n\, e^2\over 
     m\omega\tau_0}\,\sigma_*(B;\omega).
\end{equation}
Here, $\tau_0^{-1}= mv^2/\hbar^3$ is the rate of electron scattering
by the short-range potential~(\ref{potential}) in the absence of the
magnetic field. The frequency-dependent scaling field in
Eq.~(\ref{eq:scaled-B}) is related to $\omega$ and $\tau_0$ by the
expression $B_0(\omega)=\pi m c \tau_0\omega^2/2e$.

In the self-consistent Born approximation, the function
$\sigma_*(B;\omega)$ decays with the increasing magnetic field as $
B^{-1/2}$, for $B\gg B_0(\omega)$.  With the localization effects
taken into account, this dependence becomes  steeper, with
$B^{1/2}\sigma_* \propto B^{-\mu/2}$, as illustrated in
Fig.~\ref{B-dependence}.

Within the single-electron approximation, the restriction on the
magnetic field from above is imposed by the condition $\hbar\gamma\ll
k_BT$, which is equivalent to $\omega_c\ll
\tau_0\,(k_BT/\hbar)^2$, for the short-range disorder potential.
This inequality can be fulfilled simultaneously with $B\gg
B_0(\omega)$ provided $\hbar\omega \ll k_BT$. The restriction on the
magnetic field from below necessary for the system to be in the lowest
Landau level, $\hbar\omega_c\gg k_BT$, can hold for $B\sim
B_0(\omega)$ and $\hbar\omega \ll k_BT$ provided $k_BT\gg
\hbar\tau_0^{-1}$.

The latter inequality is often fulfilled for electrons on helium
surface. In this case, for $T< 0.9$~K the random potential $V({\bf
  r})$ is due mostly to capillary waves, ripplons. It has a small
correlation length and is quasistatic. For electron densities
$n\approx 0.5\times 10^8$~cm$^{-2}$ and $T=0.7$~K, the value of
$\tau_0$ is as big as $\approx 2\times 10^{-8}$~s (see
Refs.~\CITE{Lea-in-Andrei,Lea-97}). For lower $T$, the mobility which
corresponds to the effective $\tau_0\approx 10^{-7}$~s has been
observed by Shirahama {\it et al.} \cite{Shirahama-95}. The condition
$\hbar\tau_0^{-1}\ll k_BT$ can thus be easily met. Therefore the results of
the present paper fully apply to electrons on helium as long as one
can use the single-electron approximation.

The most restrictive limitation on the present theory is imposed by
many-electron effects. Even where the electrons do not form a Wigner
crystal and the 2DES is nondegenerate, these effects can determine
magnetotransport phenomena \cite{Lea-98,Dykman-79}. However, for
sufficiently strong magnetic fields, the effective coupling to a
short-range disorder potential becomes in some sense stronger than the
electron-electron interaction. Therefore the single-electron
approximation describes certain features of the magnetoconductivity
for strong fields. In particular, although the full many-electron
static magnetoconductivity is nonzero \cite{Kuehnel-00},
$\sigma_{xx}(\omega)$ can still display a peak at the frequency
(\ref{eq:peak-position}). Observation of this peak and/or its
counterpart in the magnetic field dependence of the weighted ac
conductivity $B^{1/2}\sigma_{xx}(\omega)$
(cf.~Fig.~\ref{B-dependence}) would be a clear demonstration of
single-electron localization effects in quantizing magnetic fields.

We are grateful to M.M. Fogler and S.L. Sondhi for valuable
discussions. Work at MSU was supported by the NSF through grant
no. PHY-00071059 and by the Center for Fundamental Materials
Research. L.P. was supported in part by the DOE grant
DE-FG02-90ER40542.

\appendix

\section{Guiding center formalism}
\label{app:guiding}

Here we remind the basic expressions of the guiding center formalism
needed to derive simplified expressions for the low-frequency
conductivity at a given Landau level.  Time evolution of the momentum
operator ${\bf p} = -i\hbar\nb + (|e|/c) {\bf A}$ in
Eq.~(\ref{eq:kubo-full}) is determined by the Hamiltonian
$H_0+V({\bf r})$ where $V({\bf r})$ is the
scattering potential and
\begin{displaymath}
  H_0 = \hbar\omega_c \Bigl(p_+ p_- + {1 \over 2}\Bigr).
\end{displaymath}
Here, $p_{\pm}$ are the Landau level raising and lowering operators
\begin{displaymath}
  p_\pm = (p_x \mp i  p_y)/\sqrt{2\hbar m\omega_c}, \quad  [p_-, p_+] = 1. 
\end{displaymath}
The choice of signs corresponds to $B_z=-|{\bf B}| < 0$.

The guiding center coordinates are defined as
\begin{displaymath}
 {\bf
  R}\equiv (X,Y):\quad
  X = x + { p_y \over m\omega_c},\quad
  Y = y - { p_x\over m\omega_c}.
\end{displaymath}
They commute with the momentum operator, $[{\bf R}, {\bf p}] = 0$, but
not with each other, $[X,Y]=-i l^2$.  To express the disorder
potential in terms of these variables we expand it in Fourier series,
$V({\bf r}) = \sum_{\bf q}\; V_{\bf q} \exp(i{\bf qr})$, which gives
\begin{equation}
  \label{eq:potential}
  V({\bf r}) =
  \sum_{\bf q}\; V_{\bf q} \,e^{i{\bf q}{\bf R}-q^2 l^2/4}\,
  e^{l\, q_- p_+ } \, e^{-l\, q_+p_-}, 
\end{equation}
where $ q_\pm = (q_x \mp i  q_y)/ \sqrt{2}$. 
Introducing the operators of projection to the $n$\,th Landau level $P_N$, we
can write the projected disorder potential as
\begin{eqnarray}
  \label{eq:projection-operation}
  \tilde V^{(N)}({\bf r})&\equiv& P_N\,V({\bf r}) P_N 
  = \hbar\gamma\, P_N\sum_{\bf q}\;
  \tilde V^{(N)}_{\bf q} e^{i{\bf q}{\bf R}},\\
  \tilde V^{(N)}_{\bf q}&=&(V_q/\hbar\gamma)\,L_N(q^2\,l^2/2)
  \exp(-q^2 l^2/4),  
  \label{eq:projected-potential}
\end{eqnarray}
where $L_N(z)$ is the $N\,$th Laguerre polynomial. 

Using the commutation relation $[e^{i\bf qr},p_\pm]=lq_\pm e^{i\bf
  qr}$, we can evaluate the Fourier-transformed current-current correlator in
Eq.~(\ref{eq:kubo-full}) as
\begin{eqnarray}
  \lefteqn{
  \biggl[ 1 - \Bigl({\omega \over \omega_c}\Bigr)^2 \biggr]
  \bigl\langle p_x(t)\,p_x(0) \bigr\rangle_\omega }&&\quad \nonumber \\
  &&=
 {i\omega\over \omega_c^2}
  \bigl\langle {\partial \over \partial x} V[{\bf r}(t)]\, p_x(0)
  \bigr\rangle_\omega
 -{1\over \omega_c} \bigl\langle {\partial \over \partial y}
  V[{\bf r}(t)]\, p_x(0) \bigr\rangle_\omega,
    \label{eq:correlator-momentum}
\end{eqnarray}
where we use the notation
\begin{displaymath}
  \langle p_x(t) p_x(0) \rangle_\omega =
  \int_{-\infty}^\infty \!\! dt \; e^{i\omega t}
  \langle p_x(t) p_x(0) \rangle.
\end{displaymath}
The correlators in the r.h.s. of Eq.~(\ref{eq:correlator-momentum})
can be iterated similarly, and the whole current-current correlator
can be expressed in terms of the correlators of the derivatives of the
potential $V$  (cf.\ Ref.~\CITE{Dykman-78}).

In the adiabatic limit, $\omega \ll \omega_c$, the first term in the
r.h.s.\ of Eq.~(\ref{eq:correlator-momentum}) vanishes, and from this
equation or directly from the equation of motion, for times slow
compared to $\omega_c^{-1}$, we can identify
$$
p_x(t)\rightarrow -{1\over \omega_c}{\partial \over \partial y}
  V[{\bf r}(t)]=-{i\over \omega_c}\sum_{\bf q}q_y \,V_{\bf q}\,e^{i{\bf
      qr}(t)}. 
$$
The rightmost part of this expression is proportional to the time
derivative of the guiding center coordinate, 
$$
\dot X(t)=-(l^2/\hbar)\,\sum_{\bf q}q_y \,V_{\bf q}\,e^{i{\bf
    qr}(t)}
$$
[in the right-hand sides of the last two equations one should keep only
the terms which do not contain fast-oscillating factors
$\exp(iN\omega_ct)$ with $N\neq 0$]. We can then further identify the
projected part of ${\bf
p}$ with $m {\bf \dot{R}}$, which gives Eqns.~(\ref{eq:kubo-guiding})
and (\ref{eq:kubo-formula2}).

\section{Calculation of frequency moments}
\label{app:classification}
Here we describe an efficient numerical procedure used to obtain
exact expressions of high-order diagrams in the
expansion~(\ref{eq:mom_sinus}).

We begin by rewriting Eq.~(\ref{eq:mom_sinus}) as a sum of
exponentials,
\begin{eqnarray}
\label{uncontracted}
M_{2k} &=&
-2l^2 \left( {l^2 \over 8\pi} \right)^{k+1} \sum_{{\cal C}(\left\{
    {\bf q} \right\} )} 
\sum_{\left\{ {\bf b} \right\}}
(-1)^{\sigma(\left\{ {\bf b} \right\})} \;  \nonumber \\
&&
\times
\int\! d{\bf q}_1 \ldots d{\bf q}_{2k+2}\; {\cal C}(\left\{ {\bf q}
\right\} )\,
({\bf q}_1{\bf q}_{2k+2}) \;  \nonumber \\
&&\times
\exp \left( -{l^2\over 4} \sum_{m=1}^{2k+2} q_m^2 +
i{l^2\over 4}\! \sum_{m,n=1}^{2k+2}\! \hat{B}_{mn}^{\left\{ {\bf b} \right\}}
{\bf q}_m\wedge{\bf q}_n \right).\nonumber \\
\end{eqnarray}
The inner sum is taken over all binary sequences ${\bf b} =
  (b_1,b_2,\ldots,b_{2k})$, with $b_i=0,1$. They label possible
  combinations of signs which arise from writing the sines in
  Eq.~(\ref{eq:mom_sinus}) in terms of exponentials. The quantity
  $\sigma(\left\{ {\bf b} \right\})\equiv \sum_i b_i$.  The
  antisymmetric $(2k+2)$-dimensional square matrix $\hat{B}^{\left\{
  {\bf b} \right\}}$ has the following structure:
\begin{equation}
\label{thematrix}
\hat{B}^{\left\{ {\bf b} \right\} } = \left( \begin{array}{ccccccc}
 0 & c_1 & c_2 & \multicolumn{2}{c}{\ldots} & c_{2k} & 0 \\
 -c_1 & 0 & c_2 & & & c_{2k} & 0 \\
 -c_2 & -c_2 & 0 & & & & \\
 \vdots & & & & \ddots & c_{2k} & \vdots \\
 -c_{2k} & -c_{2k} & \ldots & & -c_{2k} & 0 & \\
 0 & \multicolumn{5}{c}{\ldots} & 0 \\
\end{array} \right)\;,
\end{equation}
where $c_i = (-1)^{b_i}$.

Because of the $\delta$-functions in the contraction function ${\cal
C}(\left\{ {\bf q} \right\} )$, integration in (\ref{uncontracted})
has to be performed over $k+1$ independent wave vectors.
Up to a prefactor $({\bf q}_1\,{\bf q}_{2k+2})$, the integrand is
an exponential of the quadratic form $(l^2/2)\sum {\bf q}_i\hat
A_{ij}{\bf q}_j$, where $i,j = 1,\ldots, k+1$.  The matrix elements
$\hat A_{ij}$ are themselves $2\times 2$ matrices, $\hat A_{ij} =
-\hat I\delta_{ij} + a_{ij}\hat\sigma_y$, where $\hat\sigma_y$ is the
Pauli matrix, and $a_{ij}=-a_{ji} = 0,\,\pm 1$.

For a given contraction ${\cal C}(\{ {\bf q}\})$ and a given vector
${\bf b}$ in (\ref{uncontracted}), \ie, for the corresponding matrix
$a_{ij}$, the Gaussian integrals can be evaluated exactly, giving
\begin{equation}
\label{classifier}
I[a]=- {1\over 4^{k}} \; \prod_{i=1}^{k+1} (1+\lambda_i^2)^{-1/2}
\sum_{m=1}^{k+1} {u_{m,1}u^\ast_{m,x} \over 1+\lambda^2_m}.
\end{equation}
where $i\lambda_m$ are the eigenvalues of the antisymmetric matrices
$a_{mn}$, and $u_{m,n}$ are the components of the corresponding
eigenvectors.  The subscript $x$ takes on the value $x = k+1$ if ${\bf
q}_1$ and ${\bf q}_{2k+2}$ are independent variables in the pairing
procedure, whereas for ${\bf q}_1\!=\!-{\bf q}_{2k+2}$ we should set
$x=1$ and additionally multiply Eq.~(\ref{classifier}) by $(-1)$.

Given the large number of terms and the computational price of
calculating the exact values in each term, we did not calculate each
integral exactly.  Instead, the value of a given term (specified by
the choice of contraction ${\cal C}(\left\{ {\bf q} \right\} )$ and
the binary sequence ${\bf b}$), was calculated numerically with double
precision.  With integer-valued matrices $a_{ij}$, the number of
different values was not exceedingly large, and we used the obtained
approximate values~(\ref{classifier}) to assign each term to an
equivalence class.  The individual weights can be positive or
negative, depending on the parity of ${\sigma(\left\{ {\bf b}
  \right\})}$.  This procedure was used instead of much more tedious
manual classification of high-order diagrams.  (In addition, we
checked for several values of $k$ that the weights corresponding to
special diagrams in Fig.~\ref{fig:zero_diagram} are indeed equal to
zero).

After the classification of diagrams was completed, the exact value of
the integral in each class was obtained by taking a representative
contracted matrix ${a}_{ij}$ and calculating the Gaussian
integral~$I[a]$ algebraically.  These integrals have
rational values; the final answer for $M_{2k}$ was obtained as a
weighted sum of these rational numbers with their respective bin
weights.   As a test, we compared the results for $k=0,1,2$ with
explicit analytic calculation.

\section{Continued fraction expansion}
\label{app:continued-fraction}

The Stieltjes transform of the conductivity $\tilde\sigma_c$ is
defined by
\begin{equation}
        \label{StieltjesTrafo}
        R(z) = {1\over 2\pi\gamma} \int_{-\infty}^\infty \! d\omega \;
        {\tilde\sigma_c(\omega+\omega_c) \over  z-i\omega/\gamma}, \quad
        \re\; z > 0,
\end{equation}
while the inverse transformation has the form
\begin{equation}
        \label{inversion_formula}
        \tilde\sigma_c(\omega+\omega_c) = 2 \lim_{\varepsilon\rightarrow 0+}
        \re\,[ R(\varepsilon + i\omega/\gamma) ],
\end{equation}
The function $R$ is related to the
moments~(\ref{cyclotron_general_MOM}) by the expression
\begin{equation}
        \label{RTrafo_moments}
        R(z) = \sum_{k=0}^\infty i^k M^c_k \, z^{-k-1}.
\end{equation}

We now construct an approximation for (\ref{StieltjesTrafo}) which
applies for an even function $\tilde\sigma(\omega+\omega_c)
=\tilde\sigma(-\omega+\omega_c)$, allows for the Gaussian asymptotics
(\ref{cyclotron_asymptotic})
\begin{equation}
        \label{asymptotic_expansion}
        \lim_{\omega\rightarrow\pm\infty} {\gamma^2\over \omega^2} \ln
        \tilde\sigma_c(\omega+\omega_c) 
        = -{1\over 2\alpha}, \quad\alpha={3\over 16},
\end{equation}
and requires only a finite number of moments. 

It is known \cite{Wall-Book,Perron-Book} that an odd function $R(z)$ can
be expanded into a Jacobi-type continued fraction,
\begin{equation}
        \label{RTrafo_cfraction}
        R(z) = \Cfrac^\infty_{j=1} \biggl({\Delta_j \over
        z}\biggr),\quad \Delta_j 
        \geq 0, 
\end{equation}
where we use the notation
\begin{equation}
  \Cfrac^\infty_{j=1} \biggl({\Delta_j \over z}\biggr) \equiv
  \cfrac{1}{z + 
    \cfrac{\Delta_1}{z + 
      \cfrac{\Delta_2}{z+ \ddots}}}
\end{equation}
for the continued fraction with coefficients $\Delta_j$ and variable
$z$. The first $J$ continued-fraction coefficients
$\Delta_1,\ldots,\Delta_J$ are obtained from the normalized moments
$M^c_2,\ldots,M^c_{2J}$ by expanding the power series
(\ref{RTrafo_moments}) into the continued fraction
(\ref{RTrafo_cfraction}) using an efficient recursive
algorithm\cite{Wall-Book}. Having obtained only a finite number of
coefficients $\Delta_j$, we need to estimate the remaining ones.
Fortunately, the asymptotic behavior
(\ref{asymptotic_expansion}) implies\cite{Lubinsky-88} the following
asymptotically linear growth for the continued-fraction coefficients
\begin{equation}
        \lim_{j\rightarrow\infty} {\Delta_j \over j} = \alpha.
\end{equation}
Therefore, if the first $J$ coefficients $\Delta_1,\ldots,\Delta_J$
are found, one can then construct an approximation $R^{(J)}(z)$ to
$R(z)$ by linearly continuing $\Delta_j$ for $j>J$,
\begin{equation}
        \label{R_approx}
        R^{(J)}(z) = \Cfrac^\infty_{j=1}\biggl( {\Delta^{(J)}_j \over
        z} \biggr), 
\end{equation}
where
\begin{equation}
        \Delta^{(J)}_j = \left\{
        \begin{array}{lcl}
                \Delta_j & {\rm for} & j\leq J \\
                \Delta_J+ \alpha(j-J) & {\rm for} & j>J 
        \end{array} \right.
\end{equation}
A continuous fraction with a linearly increasing coefficient can be
written in terms of the Whittaker parabolic cylinder function $D_\nu$,
\begin{displaymath}
  T(\beta,\alpha,z)\equiv \Cfrac_{j=1}^\infty \biggl({\beta+\alpha j
    \over z}\biggr) = 
  {D_{-(\beta/\alpha)-1}(\alpha^{-1/2} z) \over
    \alpha^{1/2} D_{-\beta/\alpha}(\alpha^{-1/2} z) },  
\end{displaymath}
which is valid if $\alpha>0$, $\beta+\alpha>0$ and $\re\, z>0$, so that
we can write Eq.~(\ref{R_approx}) as
\begin{equation}
        R^{(J)}(z) = \cfrac{1}{z + \cfrac{\Delta_1}{z +
            {\displaystyle\phantom{\Delta}\atopwithdelims..%
              {\displaystyle\ddots
        {\atopwithdelims..\displaystyle z + 
        \cfrac{\Delta_{J-1}}{z + \Delta_J T(\Delta_J, \alpha,
        z)}}\!}\!}}} \; . 
\end{equation}
Applying the inversion formula (\ref{inversion_formula}) immediately
gives the restored cyclotron resonance absorption
$\tilde\sigma_c(\omega)$ as shown in
Fig.~\ref{fig:sigma_cyc_cfraction}.

\end{multicols}
\end{document}